\documentclass{an}
\usepackage{graphicx}
\usepackage{times}
\Volume{00}              
\Year{0000}              
\Month{00}               
\Pagespan{000}{000}      

\begin{document}
\lhead[\thepage]{Bachev, Strigachev: Variability of Mrk~279}
\rhead[Astron. Nachr./AN~{\bf XXX} (200X) X]{\thepage}
\headnote{Astron. Nachr./AN {\bf 32X} (200X) X, XXX--XXX}

\title{Optical Continuum Variability of the Active Galaxy Mrk~279 -- Implications for Different Accretion Regimes}

\author{R.Bachev\inst{1,2}
\and A.Strigachev\inst{1}}
\institute{
Institute of Astronomy, Bulgarian Academy of Sciences, 72
Tsarigradsko Chausse Blvd., 1784 Sofia, Bulgaria
\and
Department of Physics and Astronomy, University of Alabama, Tuscaloosa, AL 35487, USA
}

\date{Received {\it date will be inserted by the editor}; 
accepted {\it date will be inserted by the editor}} 

\abstract{
We present results from a recent broad-band monitoring in optics of the Seyfert~1 
type galaxy Mrk~279. We build and analyse the $BVRI$ light curve covering a period 
of seven years (1995 -- 2002). We also show some evidence for the existence of two 
different states in brightness and suggest, based on a modelling of the optical 
continuum, that these states may result from transition between a thin disk and an 
$ADAF$ accretion modes. We assume that the short-term variability is due to a 
reprocessing of a variable X-ray emission from an inner $ADAF$ part of the flow, 
while the long-term one may be a result from a change of the transition radius. 
Our tests show a good match with the observations for a reasonable set of accretion 
parameters, close to the expected ones for Mrk 279.
\keywords{galaxies: active -- galaxies: Seyfert -- galaxies: individual (Mrk~279) -- galaxies: photometry}
}

\correspondence{bachevr@astro.bas.bg, anton@astro.bas.bg}

\maketitle

\section{Introduction}

The continuum variability is a well-known feature of many Active Galactic Nuclei (AGNs). It is 
usually thought that the variability of radio-quiet AGNs is connected to instabilities of the 
accretion flow, feeding the central supermassive black hole, unlike the case of blazars 
where similar variations are usually attributed to processes in a relativistic jet (\cite{ul}; 
\cite{ka2}). Therefore any knowledge about the continuum variations might shed some light onto 
the accretion process and respectively -- the nature of the central engine of the active nuclei. 
Although many Seyfert galaxies are known to be variable, the variability of only a few of them 
has been studied intensively so far, which is our motivation to begin a program (\cite{ba2}) 
for optical monitoring of selected objects. 

In this paper we present the results of a broadband $B$, $V$, $R_{\rm c}$ and $I_{\rm c}$ 
monitoring of the radio-quiet active galaxy Mrk~279. Mrk~279 is a relatively bright 
$V$$\approx$$14^m$ spiral ($S0$), Seyfert~1 type galaxy, for which both continuum and emission-line 
profiles are known to vary in time (\cite{st}; \cite{sa}). Combining the results from the 
reverberation mapping, which give a broad line region (BLR) radius of about 12-17 light days 
(\cite{ma}; \cite{sa}), and the width of the broad emission lines (about 6000~$km/s$), the mass 
of the central object can be inferred -- $M_{\rm BH}\approx10^{8}~M_{\odot}$  (\cite{ho}). 
This mass and the nuclear bolometric luminosity of about 10$^{45}$~$erg/s$ give a rough estimate 
of the accretion rate of 0.01-0.1, measured in Eddington units (see also \cite{bi}). Since later 
we will propose that most probably two different accretion modes (Advection Dominated Accretion 
Flow -- $ADAF$ -- and a thin disk) operate in this object, we point out here that this rate is 
rather close to the critical accretion rate of the transition between an $ADAF$ and a thin disk 
accretion regimes, which is in general assumed to be of the same order -- 0.01-0.1 (\cite{na}).

This paper is organised as follows: observations and reductions are presented in Sect.~2; in 
Sect.~3 we present some evidence for different states in brightness; Sect.~4 compares different 
variability scenarios and shows the results of the continuum modelling, under the assumptions of 
a change of the accretion disk structure. Sect.~5 is the discussion and we present our conclusions 
in Sect.~6. The table containing the $BVR_{\rm c}I_{\rm c}$ magnitudes of Mrk~279 is given in the 
Appendix.

\section{Observations and reductions}

The major part of the observations were performed with the 0.6-m telescope of the Observatory 
of Belogradchik, Bulgaria, equipped with SBIG ST-8 CCD camera and Johnson-Cousins 
$BVR_{\rm c}I_{\rm c}$ filters (\cite{ba1}; \cite{ba2}). A few data points 
(JD~2450000+: 2066.3, 2096.3, 2099.3) were obtained using the 1.3-m telescope of the Skinakas 
Observatory, University of Crete, Greece, equipped with a Photometrics CH~360 CCD camera.

The monitoring covered a period of five years (July~1997 -- June~2002). Standard aperture 
photometry was performed in order to estimate the AGN magnitudes. Star $A$ and star $C$, whose 
magnitudes and finding charts can be found in \cite{ba2}, were used as the main standard and as a 
check respectively. The $B$-band magnitude of the main standard, not published in the paper cited 
above, was additionally calibrated -- $B=13\fm08$.

$BVR_{\rm c}I_{\rm c}$ magnitudes of Mrk~279 were measured in 66 observational epochs and are 
given in Table~1 (Appendix). The typical sampling interval was several days with the exception 
of a few larger gaps. For each observational point at least two CCD frames in each filter were 
taken, with a typical exposure time of 120~$\sec$. During the observations the seeing was usually 
2-3$\arcsec$, which presumably did not affect the aperture photometry, performed in a much larger 
diaphragm (16$\arcsec$). The standard errors of the differential photometry are typically below 
$0\fm02$ (about $0\fm05$ for $B$-band). The object showed significant variability during the 
observational period (Fig.~1) with time scales starting from 1 day. Short-term (1-4 hours) 
variations were searched extensively for more than a total of 20 hours in $V$ and $I$-bands, but 
were not detected above the limit of the photometric errors.

We used, in addition, recently published results on monitoring of Mrk~279 in optics and near-IR 
(\cite{sa}). Except spectroscopically, the object was also monitored in $BVRI$ bands with the 
1-m telescope of Wise Observatory using a 7$\arcsec$ diaphragm. Although not scaled to any 
photometric system, we were able to use these broad-band magnitudes for our purposes by fitting 
smooth curves to both data sequences and comparing these two fits in an overlapping observational 
epoch. Thus we obtained the following relations between our magnitudes and those from the Wise 
Observatory observations (\cite{sa}): \\ 
\\
$B_{\rm our}=B_{\rm Wise}+0.50$, \\
$V_{\rm our}=V_{\rm Wise}+1.38$, \\
$R_{\rm our}=R_{\rm Wise}+0.91$, \\
$I_{\rm our}=I_{\rm Wise}+0.33$, \\
\\
with a typical error of these transformations of about $0\fm03$. After rescaling the Wise data 
we built a combined light curve, covering an observational period of about seven years (Fig.~1).

\begin{figure*}[t]
\resizebox{12cm}{!}{\includegraphics*{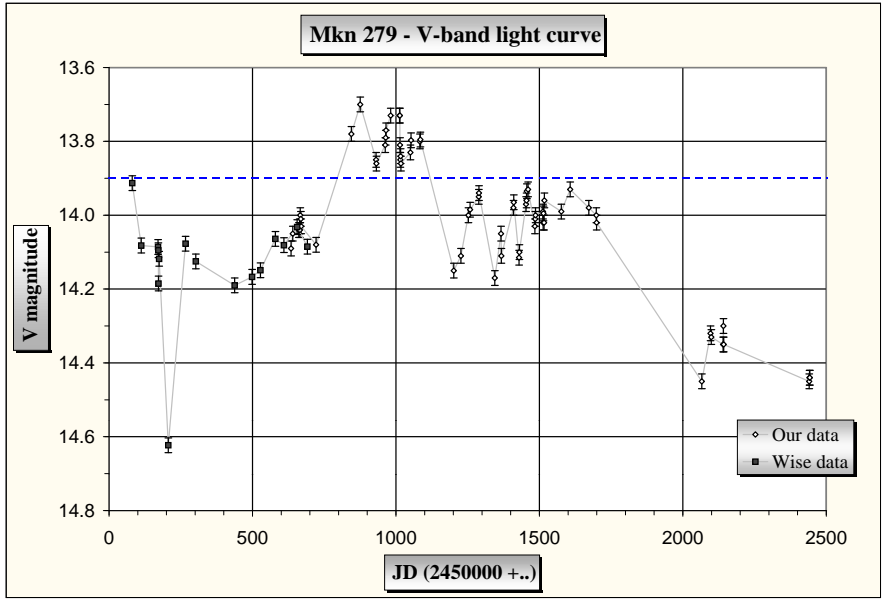}}
\caption[]{V-band light curve of Mrk~279. Magnitudes are measured in 16$\arcsec$ diaphragm. 
The filled squares are Wise Observatory magnitudes and the rombs are our data. The dashed 
line represents a possible division between two different states (see also Fig.~3).}
\end{figure*}

\section{Evidence for different states in brightness}

The $V$-band light curve of the monitored AGN is shown in Fig.~1. The variations in other bands 
occurred with no detectable time lags (less than 1 day), which, in addition to the rapid changes 
of brightness, restricts the dimension of the region producing the bulk of the variable part of 
the continuum to about 1 l.d. (or about 100~$R_{\rm G}$ in this case, $R_{\rm G} = 2GM_{\rm BH}/c^2$). 
Using the obtained light curve, we find some arguments that the object shows signatures of two 
different states in brightness. These arguments can be summarised in the following way:

\subsection{Magnitude histograms}

The distribution of the magnitude points (over 80 in total) is most likely bimodal, with a clear 
gap at $V$$\approx$$13\fm90$ where no observational points can be found (Fig.~2). Statistically, 
the presence of such a dip is quite unlikely if a $Gaussian$ distribution of the sample is assumed. 
The $Gaussian$ distribution hypothesis can be ruled out on the 95\% probability level (at least 
for $R$ and $I$-bands), as our analysis shows. 

\begin{figure*}
\resizebox{8cm}{!}{\includegraphics*{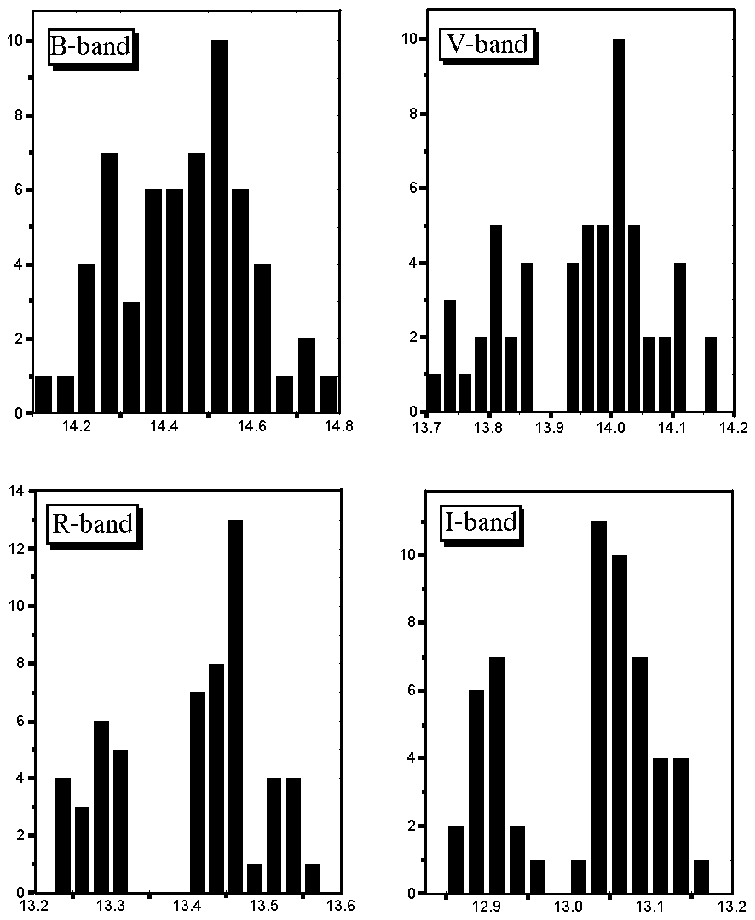}}
\caption[]{Brightness histograms for $B$, $V$, $R$ and $I$-bands respectively. The bimodal 
distribution is clearly seen (except for $B$-band). The histograms are based on both -- our 
data and the Wise Observatory data. A few observational points where the object was very faint 
are not used.}
\label{f2}
\end{figure*}

Surprisingly, the gap is best seen in $I$-band and is absent in $B$-band. This result cannot be 
attributed to the photometric errors only, which indeed increase toward shorter wavelengths. Most 
likely this gap indicates the presence of two types of variability: long-term ($\approx$~100 days) 
variations responsible for the transitions between the states, and short-term (1-10 days) variations. 
If so, the variability amplitudes must depend on the colour in different ways for the short and the 
long-term variability, producing the magnitude distribution picture shown.

\subsection{Colour-magnitude diagram}

There is a slight but well detectable jump in the colour-magnitude relation (Fig.~3) at the point 
where the transition between the states is expected, indicating that the physical processes, 
responsible for the continuum emission, probably change during that transition. Note that the 
magnitudes shown here ($V$ and $I$) cover wavelength areas, which are generally located out of the 
places where strong (and presumably variable) emission lines are present, i.e. we observe and 
analyse only the continuum variability. In fact, a similar jump is observed for the other colours 
as well, but it becomes more evident with the increase of the wavelength difference. From a 
statistical point of view, the transition process should be relatively fast, based on the absence 
of magnitude points around the jump, while the adjacent areas are heavily populated (Fig.~3).
Our interpretation is that this is possibly a manifestation of a two-state behaviour. Possible 
observational errors could hardly account for the jump (see Fig.~3 for details).

\begin{figure*}[t]
\resizebox{12cm}{!}{\includegraphics*{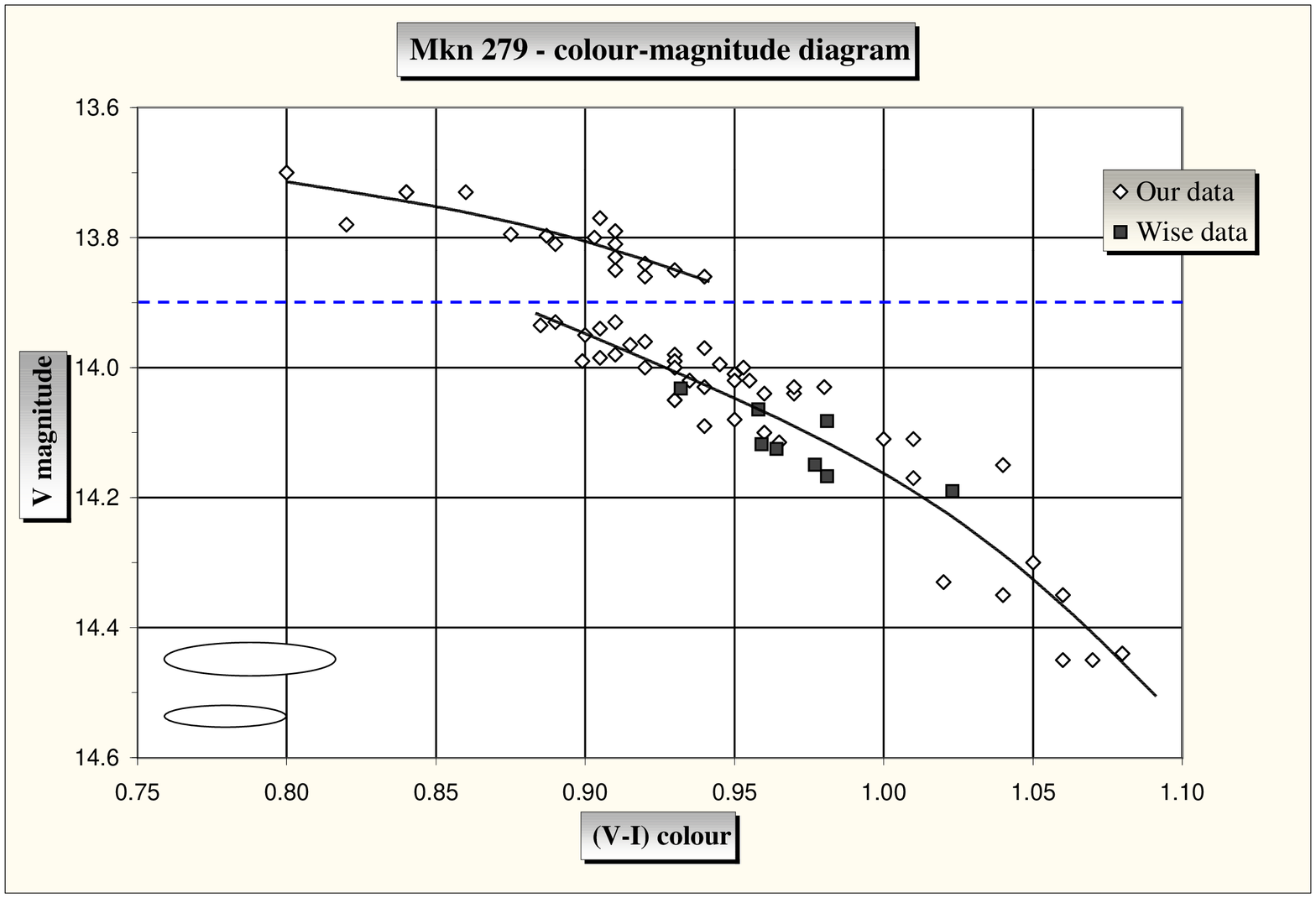}}
\caption[]{The relation between the colour ($V-I$) and the $V$-band magnitudes of Mrk~279. A jump 
in the relation is clearly seen. The thick lines are to guide the eye. The filled squares are the 
Wise Observatory magnitudes and the rombs are our data. The dashed line shows the division between 
the two states. The large ellipse at the lower left corner of the picture indicates the area that 
contains about 67\% of all $V$ vs. $V-I$ points for the check star, measured in respect to the 
main standard. As such, it it is an estimate of the photometric errors. The smaller ellipse is a 
similar prediction for the variable AGN, taking into account the much better statistics we get for 
it. It is seen that the two-state behaviour is highly unlikely to be due to photometric errors.}
\label{f3}
\end{figure*}

\subsection{Structure Function}

In order to find further arguments in favour of the two-state behaviour of Mrk~279, we invoke 
the first-order Structure Function - $SF$ (\cite{di}). This function is similar to the power 
density spectrum in some sense, but it is easier to build for an unevenly spaced data. For the 
AGN case, the $SF$ is usually characterised by a saturation time $\tau_{\rm var}$, after which the 
$SF$, initially rising with a slope of 0.3-0.7 (\cite{ka2}; \cite{co}), begins to turn over and 
remains nearly a constant. At timescales grater than the saturation time, $\tau_{\rm var}$, the 
amplitude of the variations does not increase, indicating that $\tau_{\rm var}$ is probably 
associated with some of the timescales of the physical processes driving the variability.

In order to search for differences in the nature of the short-term (1-10 days) and the long-term 
variability ($\approx$~100 days), we build $SF$'s separately for the lower and the higher state. 
These functions are based on 30 and 18 observational points respectively (Fig.~4) and cover the 
periods where the object has been most intensively monitored. Although there is a significant 
scatter, we see indications that the nature of the variability may really be different for these 
two states. In particular, the saturation time ($\tau_{\rm var}$) appears to be much shorter for 
the higher state (about 5 days) than for the lower state (about 20 days), Fig.~4.

\begin{figure*}[t]
\resizebox{8cm}{!}{\includegraphics*{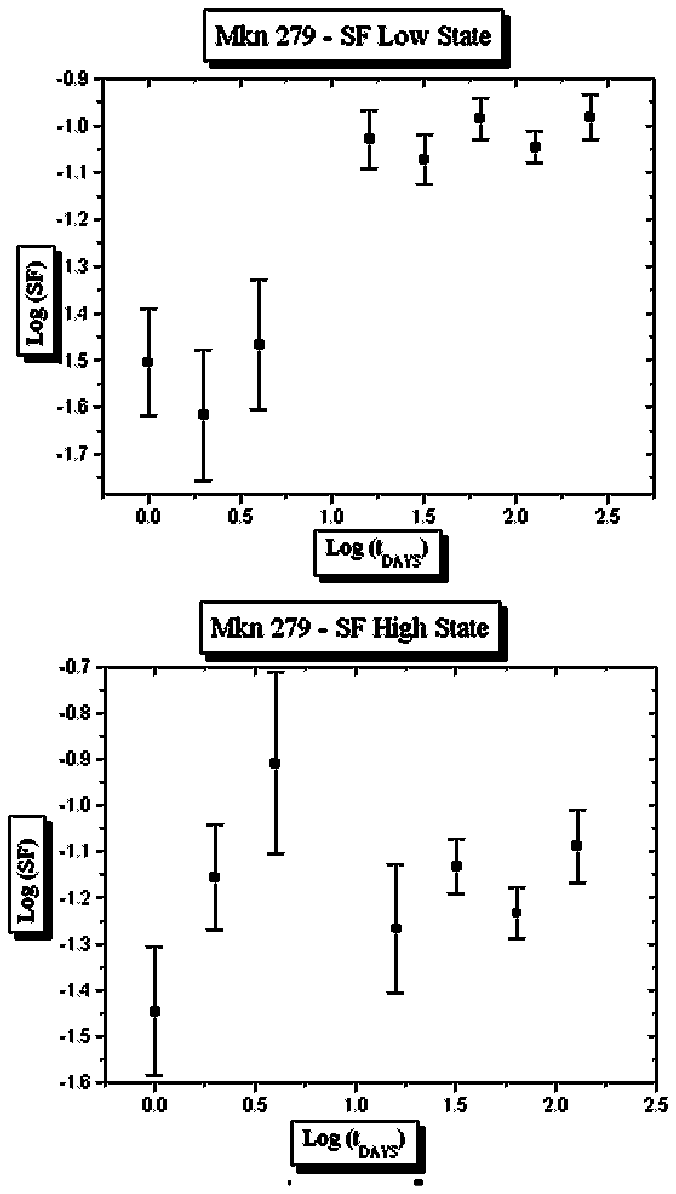}}
\caption[]{The $SF$s for the upper state (based on 18 points) and the lower state (based on 30 points, 
starting immediately after the high state). It is important to note that the saturation time scale is 
very short, in order of several days. Furthermore this time is different for the lower and the higher 
states -- resp. 20 and 3-5 days.}
\label{f4}
\end{figure*}

\section{Modelling the continuum variability}

In this section we consider several schemes that can, in principle, account for the observed 
variability picture and elaborate the possibility that a change of the accretion disk structure 
is primarily responsible for the bimodal behaviour.

\subsection{Accretion disk structure changes}

We adopted a simple model to test the possibility of reproducing the observed variability. The 
accretion flow is assumed to consist of an inner advection part operating at $r < R_{\rm tr}$ and 
an outer, thin-disk part at $R_{\rm tr} < r < R_{\rm out}$. The thin disk emits as a blackbody in 
optics. The inner optically thin ADAF is assumed not to contribute to the optical continuum itself, 
but its hard X-rays might irradiate the outer part, increasing the temperature and producing some 
extra optical flux there. Since the short-term variations occur on time-scales of about a day, we 
accept that they are produced by variations of the central X-ray emission, reprocessed by the outer 
parts in optics. The geometry of the central X-ray emitting region is taken, in our simple model, 
to be a uniform sphere with radius $R_{\rm tr}$. In such a case, the bulk of the emission will come 
from a typical height $H_{\rm s}=4R_{\rm tr}/3\pi$ above the disk. A more realistic geometry might 
be more relevant, but we believe it will not change the results much. As one can show, the absorbed 
X-ray flux at a given radial distance ($F_{\rm x}$) is roughly proportional to the product of 
$H_{\rm s}$ and $L_{\rm x}$, and we consider $L_{\rm x}$ as a free parameter anyway (see below). 
Furthermore, any specific geometry could hardly be justified, so we adopt the simplest case. Any 
flaring and possible warping of the disk were neglected. We assume that the switch between the 
states is due to a relatively fast change of the transition radius between two metastable positions 
($R_{1}$ and $R_{2}$), which can occur for some critical accretion rate - $\dot{m}_{\rm cr}$ 
(see below).

We take $R_{\rm tr}$, $R_{\rm out}$, $L_{\rm x}$, $M_{\rm BH}$ and $\dot{m}$ to be free parameters 
($\dot{m}$ is the accretion rate expressed in Eddington units). We run a series of tests changing 
these parameters within some reasonable limits ($log(M_{\rm BH})=6-9$, $\dot{m}=0.001-1$, 
$R_{\rm tr}=3-200R_{\rm G}$). Our goal is to find a set of parameters that will reproduce our data 
well. To test our scheme we have to find out if the following will be fulfilled:

\begin{enumerate}

\item The adopted scheme could in principle reproduce the observed variability -- time-scales, colour 
trends, jumps, etc. only by changing $R_{\rm tr}$ and $L_{\rm x}$ for given $M_{\rm BH}$ and $\dot{m}$.

\item The best working set of parameters found will be close to those already adopted for this object 
(see Sect.~1).

\item These parameters will be non-contradictory in between and consistent with the theoretical predictions.

\end{enumerate}

In other words we try to find out if a non-contradictory accretion disk structure can account for the 
observed optical variability.

The effective temperature of the disk is 
$T(r) = T_{\rm G}(r)[1+F_{\rm x}(r)/({\sigma}T_{\rm G}(r)^4)]^{1/4}$ K.
This is the standard thin disk effective temperature, 
$T_{\rm G}(r) = 3.5\,10^7[\dot{m}/({\eta}M_{\rm BH}r^3)]^{1/4}$ K (\cite{fr}),
corrected for the absorbed X-ray flux coming from the central $ADAF$. 
Here the radial distance $r$ is expressed in $R_{\rm G}$; the accretion efficiency is taken ${\eta}=0.1$. 
Thus, by assuming a blackbody radiation, we get for the overall emission at frequency $\nu$ to be 
$F_{\nu} {\propto} \int_{R_{\rm tr}}^{R_{\rm out}} B_{\nu}(r)rdr$; $B_{\nu}(r)$ is the Planck function. This 
flux can be consequently converted into magnitudes in order to be compared with the real observations.

Although one may think that there are many different sets of parameters that would be able to 
reproduce the observations equally well, the results from our tests indicate otherwise. For instance, 
to reproduce the overall colour-magnitude trend we had to restrict $log(M_{\rm BH}) > 7$ and $\dot{m} < 0.1$ 
for any reasonable $R_{\rm tr}$; otherwise, the slope turned out to be too steep, meaning an almost 
non-chromatic variability. We have to say that, in general, we failed to reproduce the colour jump in the way 
it was observed for any reasonable values of the parameters and $R_{\rm out} = \infty$. The lower 
(corresponding presumably to a large $R_{\rm tr}$) part of the colour-magnitude relation, appeared to be 
shifted to the red in respect to the upper part, which is exactly the opposite of what we see. We succeeded, 
however to get approximately the "right" jump pattern by reducing significantly $R_{\rm out}$ to 
100 - 120 $R_{\rm G}$. As we will see in the next section, such a low value for the outer boundary of the 
accretion disk is not entirely unrealistic from a theoretical point of view. 

Fig.~5 shows two solutions for $\dot{m} = 0.01$ and $log(M_{\rm BH})$ = 7.5 and 8 respectively. The central 
X-ray flux $L_{\rm x}$, responsible for the short-term variations, is still a free parameter, but is limited in 
a way that will not allow the absorbed (and re-emitted) flux to dominate in the total optical flux from the 
disk. The $R_{\rm tr}$ changes between $R_{1} = 20R_{\rm G}$ and $R_{2} = 50R_{\rm G}$ during the transition 
between the states. This choice is an empirical one, results from many tests, and provides a maximal separation 
between the states (see also Fig.~3), reaching nearly the observed one, without requiring unrealistically 
high $R_{2}$. Other choices either would show no significant discontinuity in the colour-magnitude relation 
or a relatively much redder lower part, contrary to what was observed. Smaller values of $R_{1}$ seem to work 
somewhat worse but still acceptably well.

\begin{figure*}[t]
\resizebox{8cm}{!}{\includegraphics*{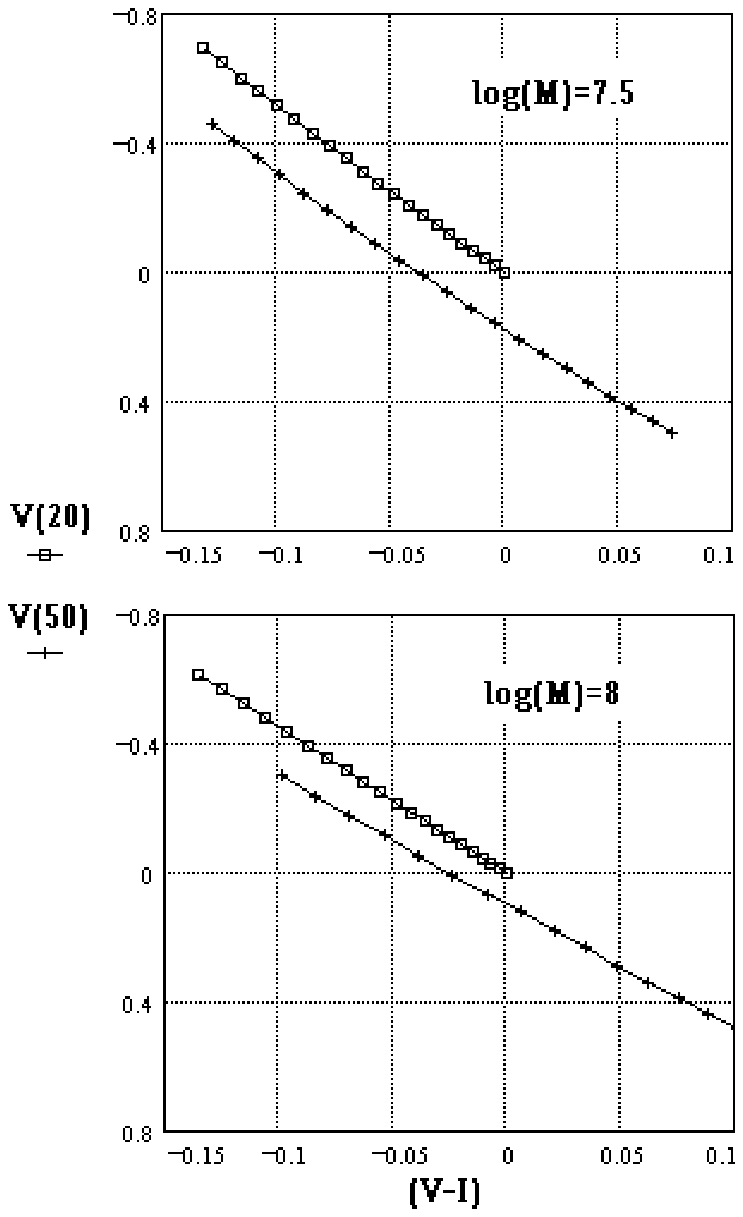}}
\caption[]{Results from the continuum variations modelling -- $(V-I)$ colour vs. $V$-magnitude 
(relative units used). Here $\dot{m}=0.01$, $R_{\rm out}=100R_{\rm G}$ and $L_{\rm x}$ is a free parameter, 
which changes produce the short-term optical variability (the thick lines). Each point along the curves 
represents a change of $log(L_{\rm x})$ by 0.05, the full range of change of $L_{\rm x}$ is the same for both 
curves. The two curves shown are for different $R_{\rm tr}$ -- 20 (the upper one) and 50 (the lower one) 
$R_{\rm G}$. The upper picture is for a black hole mass $log(M_{\rm BH})$ = 7.5, the lower one -- 
$log(M_{\rm BH})$ = 8. These two solutions are found to mimic best the true variability picture (Fig.~3).}
\label{f5}
\end{figure*}

In other words we find that these two parameter sets (see above, also Fig.~5) reproduce the variability 
pattern best and are among the few that work at all.

Let us now see if the adopted set of parameters is physically relevant and intrinsically non-contradictory. 
First, one notices that the $M_{\rm BH}$ and $\dot{m}$ values are indeed quite close to what we expected 
(Sect.~1). We also note that the saturation times (Sect.~3.3) are in very good agreement with the $Keplerian$ 
times (or free-fall times) for about 20 and 50 $R_{\rm G}$ -- respectively 4.5 and 18 days for 
$log(M_{\rm BH}) = 8$. The significance of this fact is discussed in the next section.

In addition, to be able to account for a transition between the states, the accretion rate has to be close to 
some critical value $\dot{m}_{\rm cr}$, where an intermittent transition between the accretion modes would 
eventually occur. In fact, most theoretical approaches (\cite{es}, \cite{ro}) predict a discontinuous change 
of the transition radius for some certain (critical) accretion rate. In their model, R\'{o}zanska \& Czerny 
(2000) find $\dot{m}_{\rm cr} {\approx} 0.07(\alpha_{0.1})^{3.3}$, and we adopt their result, but other models 
might have different $\dot{m}_{\rm cr}$ (see the same paper for a comparison). Due to the strong dependence on 
the viscosity parameter ($\alpha_{0.1}={\alpha}/0.1$), one can easily get $\dot{m}_{\rm cr}=0.01$, without 
requiring unreasonable values of $\alpha$ (${\alpha}=0.05$ in that case). The same model predicts (for such 
$\dot{m}$ and $\alpha$) a discontinuous change of $R_{\rm tr}$ between $3R_{\rm G}$ and ${\sim}25R_{\rm G}$. 
These values are slightly different from what we find (20$R_{\rm G}$ and 50$R_{\rm G}$ resp.), but a perfect 
match can hardly be expected since neither of the models producing these results could claim perfection.  

\subsection{Microlensing}

Another reasonable possibility that could have a chance to explain the complex variability picture is a 
microlensing event that produces a discontinuity in the light curve. Within this explanation, the colour 
changes are not entirely unexpected since only a part of the disk might be magnified (\cite{yo}). The duration 
of the upper state (about a year) is not unreasonable as well. The microlensing model faces some difficulties, 
however. It is hard to explain, for instance, the change of the saturation time (Sect.~3.3). One would also 
expect a gradual rise and fall of the light curve, while we see quite "flat" upper state (Fig.~1). Furthermore, 
the colour changes have to be more gradual as the lensing body passes over different parts of the disk (the 
disk temperature changes gradually), but this is not the case here. Although a microlensing event cannot be 
entirely ruled out, all the difficulties mentioned above seem to favour the former, accretion disk structure 
changes possibility.   

\subsection{Star disruption}

Star disruption near the black hole has been invoked sometimes to account for the feeding of the central engine. 
Recent detailed models indicate however, that such an event should be rare ($10^{-4 \div -6} yr^{-1}$) and cannot 
be the primary source of accreting gas (see \cite{ulm} for a review). Still it might, in principle, take place 
and eventually account for the two-state character of the light curve of Mrk 279. However, this scenario, as the 
previous one, encounters significant difficulties that make it unlikely:

\begin{enumerate}

\item The rather flat upper state is not expected. Instead, the brightness should decrease as a power low 
(~ $t^{-5/3}$), as the models show. 

\item The colour changes should not be as they were observed. The brightness temperature of a standard thin disk 
is about 3-5000 K (assuming $F_{\nu} {\propto} {\nu}^{1/3}$) for the optical -- near IR region, while the 
accreting disrupted material will have a temperature of about $10^5$ K or more (\cite{ulm}). Therefore the 
upper state should appear relatively bluer not relatively redder in respect to the lower state, which is 
contrary to what the observations show.

\item In addition to the fact that this event is found to be very rare, it is worth mentioning that the black hole 
of such a mass ($10^8 M_{\odot}$ -- Sect.~1), would rather swallow a Solar type star than disrupt it, i.e. 
somewhat lower $M_{\rm BH}$ is needed for the process to be effective. 

\item Again, as in the previous case, the change of the saturation time will hardly be explained.

\end{enumerate}

\section{Discussion}

Although none of the arguments presented in Sect.~2 alone can be convincing enough, altogether they give us 
confidence to conclude that the two states mentioned previously are probably real, and are not due to a small 
number of observations, photometric errors or just serendipity. Such a two-state variability has not been 
reported for the AGNs so far, however most of them have not been monitored long enough in different colours. 
We also note an interesting $X$-ray analogy, reported by \cite{lu}. They find an 
{\it "Ionising luminosity -- X-ray spectral index"} relation for many $X$-ray sources, which is remarkably 
similar to our optical colour--magnitude diagram (Fig.~3) built for one but variable source. \cite{lu} 
interpret their result in terms of the thin disk -- $ADAF$ paradigm, assuming that for some certain accretion 
rate the thin disk mode switches to an $ADAF$ mode. In their work the accretion rate value, represented by 
the ionising luminosity, is assumed to determine the accretion regime. Similarly, we think that the best 
explanation for the observed two-state behaviour of Mrk~279 should be searched in a change of the transition 
radius ($R_{\rm tr}$) between an inner $ADAF$ (or any other type low-radiative solution - $CDAF$, $ADIOS$) 
and an outer thin disk (the radius moves inward from the lower to the higher state).

The exact variability pattern can hardly be reproduced by any simple theoretical scheme. Nonetheless, we have 
shown (Sect.~4) that an accretion disk structure, consisting of an inner X-ray variable $ADAF$ section, and an 
outer thin disk with variable transition radius can reproduce the variability pattern well -- both the short 
and the long term variations, as well as the colour changes. In order to be successful, such a scheme imposes 
certain restrictions on the main governing parameters -- $M_{\rm BH}$, $\dot{m}$, $R_{\rm tr}$, $R_{\rm out}$, 
$\alpha$. We found that all these parameters are in a good agreement in between, according to the theoretical 
predictions. The assumption of a variable $R_{\rm tr}$ is not unusual. It is actually required for NGC~5548 
as suggested by \cite{chi}, who performed very similar modelling of the continuum of that object.

The only clear problem with the scheme we propose is that the outer thin disk has to be truncated closer to 
the centre than one would normally expect, i.e. at about $100R_{\rm G}$. However such a small-scale disk is 
not necessarely irrelevant. One reason is that a thin disk is a subject of selfgravitation in its outer parts 
(see for details \cite{col2}, \cite{col1}). The selfgravitating part of the disk will eventually transform 
into distinct selfgraviting clouds, which if irradiated from the centre, will produce line emission rather 
than continuum. It has been proposed, indeed, that the seflgraviting part of a thin disk is the place where 
the broad lines originate. Some recent results actually suggest that the BLR should exist at a distance where 
an optically thick disk is already not present. \cite{rok}, for instance, find that sources pointed almost 
face on with respect to the observer will still reveal $broad$ $non$-$shifted$ emission line profiles. 
This fact rules out the possibility that the major part of the emission is coming from a disk or any structured 
flow (inflow, outflow) above an optically $thick$ disk. The disk is ruled out since the profiles are too broad 
for a face-on orientation; the inflow/outflow is ruled out since we will not be able to see the far side of 
the region and the profiles will appear shifted. The only remaining possibility is a motion of clouds, with 
probably a flattened distribution, at a place where the disk does not extend; otherwise the clouds will strike 
the disk and will likely be destroyed. A good candidate for such a region is indeed the selfgravitating outer 
region of a thin disk. The significant breadth of the lines as well as the very broad line wings often observed 
suggest that such a region could start as close as 100 $R_{\rm G}$. Note that, if this is correct, the line 
width will represent mostly the vertical velocity dispersion for a face-on orientation, rather than the 
$Keplerian$ velocity, which would lead to underestimates of the black hole masses based on the broad line 
velocity.

Let us consider the results from recent numerical computations for the radius at which the disk becomes 
selfgraviting -- $R_{\rm SG}$, summarised by Collin (2001). They find $R_{\rm SG}\simeq500 R_{\rm G}$ for 
$10^8 M_{\odot}$ and $\dot{m}=0.01$, but the transition radius actually gets smaller when $\alpha$ is 
less than the canonical value of 0.1, which seems to hold here, based on the requirement 
$\dot{m}_{\rm cr}\simeq 0.01$ (see Sect. 4.1). Other uncertainties, such as the exact value of the 
Toomre parameter $Q$, can also alter (and eventually further reduce) $R_{\rm SG}$, making it quite 
close to the value we get for $R_{\rm out}$. 

We would also like to emphasise the connection between the $Keplerian$ times at $R_{1}$ and $R_{2}$, 
and the saturation times. Note that even though the brightness can change with a time interval as short 
as 1 day, the saturation of the variability amplitude may require a much longer time, close to 
$\tau_{\rm var}$ (Sect.~3.3). It is tempting to connect the saturation time with instabilities of some 
sort taking place and producing variable X-ray emission at $R_{\rm tr}$, and therefore being naturally 
associated with the dynamical timescale there. They can either be some quasi-periodic oscillations with 
the $Keplerian$ frequency (\cite{gar}), or "hot spots" of some sort originating at $R_{\rm tr}$ and 
falling inside with an almost free-fall velocity (close to the radial velocity of an $ADAF$) as 
described by \cite{ka2} in their disk-instability model. In both cases the saturation should occur at 
a timescale close to the dynamical one at $R_{\rm tr}$.

In addition, it has to be pointed out that no stable solution providing a transition between an outer 
thin disk, and an inner $ADAF$ has been found so far, what indirectly suggests that instabilities of 
some sort might occur at the transition radius, as we suggest based on our findings. Due to a scarce 
sampling for the object we monitored, we can estimate only the upper limit for the duration of the 
transition process. It is about 100 days but in fact, the real transition can be much shorter. Most 
probably this time should be associated with the thermal time scale of a thin disk (see \cite{du} for 
a review). Although it is not clear what triggers the transition process and switches between the 
regimes, it can be speculated that some small changes of accretion rate about the critical value can 
produce the observed picture. In fact, the estimated accretion rate of Mrk~279 is rather close to that 
critical value as we mentioned above. The transition between the accretion regimes, connected with a 
jump in accretion efficiency, can result in the dip in the brightness histogram that we observe 
(Fig.~2). 

An ultimate test to the model we propose to account for the overall picture of variability of Mrk~297 
could be provided by a hard X-ray monitoring, if such were performed. Our results suggest on average 
about 2-5 times higher X-ray flux during the lower $ADAF$ state (for similar $V$-band magnitudes).

\section{Conclusions}
We describe the nature of Mrk~279's optical variability in the following way: the long-term variations 
(100-300 days) are dominated by the transition between the states. These variations are probably 
caused by a change of the transition radius between an (inner) $ADAF$ and (outer) thin disk state 
which is most likely due to a small change of the accretion rate around some critical value. The higher 
state is observed when the transition radius moves inward, increasing the area of the energetically 
more efficient outer thin disk. The short-term variability could be attributed to different sources. 
We suppose that it is connected to the processes (instabilities) that take place around the transition 
radius, resulting in smaller variability time-scale for the higher state. 

Other explanations, like gravitational lensing or star disruption near the central black hole, can in 
principle also account for the observations, but these scenarios face significant difficulties to 
match well the variability picture. Additional EUV/X-ray data, if were available during the optical 
monitoring, can clarify the role of transition radius changes in cases like Mrk~279. 

We summarise the results of our work in the following way:

\begin{enumerate}

\item Using our observational data and data from the literature, we build seven-years optical 
$BVRI$ light curve of the Seyfert 1 type galaxy Mrk~279. The typical errors of the photometry 
are about $0\fm02$.

\item Analysing the data we find arguments in favour of the possibility that Mrk~279 shows 
different states of brightness. They are characterised by different colour-magnitude relations 
and different short-term variability. 

\item We find that these states may result from a transition between the thin disk and the 
$ADAF$ accretion modes, as our modelling shows. This hypothesis does not confront the 
observational data.

\end{enumerate}

Finally, we would like to emphasise that such regular multicolour observations, even performed 
with the facilities of smaller observatories, can bring important information about the AGN 
variability. The physics of accretion flows is not yet well understood, and we think that such 
observations might help when the processes, taking place at the AGN centres, are modelled.

\acknowledgements
CCD ST-8 at the Observatory of Belogradchik is provided by Alexander von Humboldt foundation, Germany. 
Skinakas Observatory is a collaborate project of the University of Crete, the Foundation for Research 
and Technology - Hellas, and the Max-Planck-Institut f\"ur Extraterrestrische Physik. 
The authors are greatful to Dr. Becky Grouchy for some technical help, as well as to our 
anonymous referee for his/her criticism that helped to improve much this paper.

\begin{appendix}
\section{Table}
\begin{table}[h]
\caption{Magnitudes of Mrk~279 (our observations only)}
\begin{tabular}{rcccc}
\hline
JD (2450000+) & B & V & R & I         \\
\hline
 634.4 & 14.6 & 14.09 & 13.52 & 13.15 \\
 640.5 & 14.6 & 14.05 & 13.50 & 13.12 \\
 661.4 & 14.5 & 14.04 & 13.46 & 13.08 \\
 664.4 & 14.4 & 14.03 & 13.45 & 13.05 \\
 665.4 & 14.5 & 14.04 & 13.45 & 13.07 \\
 666.5 & 14.5 & 14.00 & 13.45 & 13.07 \\
 668.4 & 14.6 & 14.01 & 13.44 & 13.06 \\
 669.4 & 14.5 & 14.03 & 13.44 & 13.06 \\
 721.3 & 14.5 & 14.08 & 13.51 & 13.13 \\
 844.6 & 14.2 & 13.78 & 13.31 & 12.96 \\
 875.6 & 14.1 & 13.70 & 13.24 & 12.90 \\
 930.4 & 14.3 & 13.85 & 13.32 & 12.94 \\
 931.5 & 14.4 & 13.86 & 13.32 & 12.94 \\
 962.5 & 14.2 & 13.81 & 13.29 & 12.90 \\
 964.5 & 14.3 & 13.79 & 13.24 & 12.88 \\
 965.5 & 14.3 & 13.77 & 13.24 & 12.87 \\
 981.5 & 14.2 & 13.73 & 13.25 & 12.89 \\
1012.3 & 14.3 & 13.73 & 13.25 & 12.89 \\
1013.3 & 14.3 & 13.73 & 13.24 & 12.87 \\
1014.3 & 14.3 & 13.81 & 13.29 & 12.92 \\
1015.3 & 14.3 & 13.85 & 13.29 & 12.92 \\
1016.3 & 14.3 & 13.84 & 13.31 & 12.92 \\
1017.3 & 14.3 & 13.86 & 13.30 & 12.92 \\
1050.5 & 14.4 & 13.83 & 13.28 & 12.92 \\
1052.3 & 14.2 & 13.80 & 13.29 & 12.91 \\
1083.4 & 14.3 & 13.80 & 13.26 & 12.90 \\
1085.3 & 14.2 & 13.80 & 13.29 & 12.92 \\
1201.5 & 14.8 & 14.15 & 13.53 & 13.11 \\
1226.5 & 14.6 & 14.11 & 13.53 & 13.11 \\
1252.6 & 14.5 & 14.00 & 13.47 & 13.08 \\
1258.5 & 14.4 & 13.99 & 13.46 & 13.08 \\
1288.5 & 14.4 & 13.95 & 13.42 & 13.05 \\
1289.4 & 14.4 & 13.94 & 13.41 & 13.04 \\
1344.4 & 14.6 & 14.17 & 13.56 & 13.16 \\
1366.4 & 14.5 & 14.05 & 13.50 & 13.12 \\
1367.3 & 14.7 & 14.11 & 13.48 & 13.10 \\
1409.3 & 14.4 & 13.98 & 13.44 & 13.05 \\
1410.4 & 14.4 & 13.97 & 13.41 & 13.05 \\
\hline
\end{tabular}
\end{table}

\setcounter{table}{0}
\begin{table}[h]
\caption{\it continued}
\begin{tabular}{rcccc}
JD (2450000+) & B & V & R & I         \\
\hline

1429.3 & 14.7 & 14.12 & 13.54 & 13.15 \\
1430.3 & 14.6 & 14.10 & 13.53 & 13.14 \\
1453.2 & 14.6 & 13.97 & 13.43 & 13.03 \\
1455.3 & 14.4 & 13.96 & 13.42 & 13.04 \\
1456.3 & 14.5 & 13.94 & 13.42 & 13.05 \\
1460.2 & 14.4 & 13.93 & 13.40 & 13.02 \\
1484.3 & 14.5 & 14.03 & 13.46 & 13.09 \\
1485.2 & 14.5 & 14.01 & 13.47 & 13.06 \\
1486.2 & 14.5 & 14.00 & 13.44 & 13.05 \\
1512.2 & 14.5 & 13.99 & 13.44 & 13.06 \\
1513.2 & 14.6 & 14.02 & 13.47 & 13.07 \\
1514.2 & 14.4 & 14.00 & 13.43 & 13.05 \\
1516.2 & 14.6 & 14.02 & 13.46 & 13.07 \\
1517.2 & 14.5 & 13.96 &  --   &  --   \\
1576.6 & 14.5 & 13.99 & 13.46 & 13.09 \\
1607.5 & 14.4 & 13.93 & 13.42 & 13.04 \\
1672.5 & 14.5 & 13.98 & 13.44 & 13.07 \\
1698.3 & 14.5 & 14.00 & 13.45 & 13.07 \\
1699.3 & 14.5 & 14.02 & 13.47 & 13.09 \\
2066.3 & --   & 14.45 & 13.93 & 13.39 \\
2096.3 & --   & 14.32 & 13.78 & --    \\
2099.3 & --   & 14.33 & 13.78 & 13.31 \\
2140.4 & 14.9 & 14.35 & 13.73 & 13.31 \\
2141.4 & 14.7 & 14.30 & 13.68 & 13.25 \\
2142.4 & 14.6 & 14.35 & 13.70 & 13.29 \\
2440.4 & 15.0 & 14.45 & 13.85 & 13.38 \\
2441.4 & 14.9 & 14.44 & 13.83 & 13.36 \\
2442.4 & 15.0 & 14.44 & 13.81 & 13.36 \\
\hline
\end{tabular}
\end{table}

\end{appendix}


\begin{thebibliography}{}

\bibitem[Bachev et al. 1999]{ba1}  Bachev, R., Strigachev, A., Petrov, G.T., et al.: 1999, Bulgarian 
        Journal of Physics 5/6, 1
\bibitem[Bachev et al. 2000]{ba2} Bachev, R., Strigachev, A., Dimitrov, V.: 2000, A\&AS 147, 175
\bibitem[Bian \& Zhao 2003]{bi} Bian, W., Zhao Y.: 2003, MNRAS 343, 164 
\bibitem[Chiang \& Blaes (2003)]{chi} Chiang, J., Blaes, O.: 2003, ApJ 586, 97
\bibitem[Collier \& Peterson 2001]{co} Collier, S., Peterson, B.M.: 2001, ApJ 555, 775
\bibitem[Collin 2001]{col1} Collin, S., in "Advanced Lectures on the Starburst-AGN Connection", Puebla 
        Mexico, World Scientific: 2001, p.167 
\bibitem[Collin \& Hur\'{e} 2001]{col2} Collin, S., Hur\'{e}, J.-M.: 2001, A\&A 371, 50
\bibitem[di Clemente et al. 1996]{di} di Clemente, A., Giallongo, E., Natali., G., et al.: 1996, ApJ 463, 466 
\bibitem[Dubus 2002]{du} Dubus, C.: 2002, astro-ph/0206218
\bibitem[Esin et al. 1997]{es} Esin, A.A., McClintock, J.E., Narayan R.: 1997, ApJ 489, 865
\bibitem[Frank et al. 2002]{fr} Frank, J., King, A., Raine, D.: 2002, "Accretion Power in Astrophysics", 
        Cambridge University 
\bibitem[Gracia et al. 2003]{gar} Gracia, J., Peitz, J., Keller, C., Camenzind, M.: 2003, MNRAS 344, 468
\bibitem[Ho 1998]{ho} Ho, L.C.: 1998, in "Observational Evidence for Black Holes in the Universe", Kluwer,  
        Ed. S.K. Chakrabarti
\bibitem[Kawaguchi et al. 1998]{ka1} Kawaguchi, T., Mineshige, S., Umemura, M., et al.: 1998, ApJ 504, 671
\bibitem[Kawaguchi \& Mineshige 1999]{ka2} Kawaguchi, T., Mineshige, S.: 1999, in "Active Galactic Nuclei 
        and Related Phenomena", IAUS 194, p.356
\bibitem[Lu \& Yu (1999)]{lu} Lu, Y., Yu, Q.: 1999, ApJ 526, L5
\bibitem[Maoz et al. 1990]{ma} Maoz, D., Netzer, H., Leibowitz, E., et al.: 1990, ApJ 351, 75
\bibitem[Narayan et al. 1998]{na} Narayan, R., Mahadevan, R., Quataert, E.: 1998, in "Theory of Black Hole 
        Accretion Discs", Eds. M.A. Abramowicz, G. Bjornsson and J.E. Pringle, Cambridge, p.148
\bibitem[Rokaki et al. (2003)]{rok} Rokaki, E., Lawrence, A., Economou, F. et al.: 2003, astro-ph/0301405
\bibitem[R\'{o}zanska \& Czerny 2000]{ro} R\'{o}zanska, A., Czerny, B.: 2000, A\&A 360, 1170
\bibitem[Santos-Lleo et al. 2001]{sa} Santos-Lleo, M., Clavel, J., Schulz, B., et al.: 2001, astro-ph/0102356
\bibitem[Stirpe 1991]{st} Stirpe, J.: 1991, in "Variability of Active Galaxies", Eds. W.J. Duschl, S.J. Wagner, 
        M. Camenzind, Springer-Verlag, p.71.
\bibitem[Ulmer 1999]{ulm} Ulmer, A.: 1999, ApJ 514, 180
\bibitem[Ulrich et al. 1997]{ul} Ulrich, M.-H., Maraschi, L., Urri, M.C.: 1997, ARA\&A 35, 445
\bibitem[Yonehara et al. 1999]{yo} Yonehara, A., Mineshige, S., Fukue, J. et al.: 1999, A\&A 343, 41

\end{thebibliography}
\end{document}